\documentclass[10pt,journal]{IEEEtran}

\IEEEoverridecommandlockouts
\usepackage{cite}
\usepackage{amsmath,amssymb,amsfonts}
\usepackage{graphicx}
\usepackage{textcomp}
\usepackage{multicol}
\usepackage{algorithm}
\usepackage{algpseudocode}
\usepackage{amsmath}
\usepackage{graphics}
\usepackage{epsfig}
\usepackage{epstopdf}
\usepackage{xcolor}
\def\BibTeX{{\rm B\kern-.05em{\sc i\kern-.025em b}\kern-.08em
    T\kern-.1667em\lower.7ex\hbox{E}\kern-.125emX}}
\begin{document}

\title{DeepEDN: A Deep Learning-based Image Encryption and Decryption Network for Internet of Medical Things}

\author{Yi Ding, ~\IEEEmembership{Member,~IEEE}, Guozheng Wu, Dajiang Chen, ~\IEEEmembership{Member,~IEEE}, Ning Zhang, ~\IEEEmembership{Senior Member,~IEEE}, Linpeng Gong, Mingsheng Cao, ~\IEEEmembership{Member,~IEEE}, and Zhiguang Qin, ~\IEEEmembership{Member,~IEEE}
\thanks {Corresponding authors: Dajiang Chen (djchen@uestc.edu.cn) and Ning Zhang (ning.zhang@tamucc.edu)}
\IEEEcompsocitemizethanks{
\IEEEcompsocthanksitem{This work is jointly supported by NSFC (No. 61872059 and 61502085), Natural Science Foundation of Guangdong Province (No. 2018A030313354) and China Postdoctoral Science Foundation funded project (No. 2015M570775).}
\IEEEcompsocthanksitem Yi Ding is with the Network and Data Security Key Laboratory of Sichuan Province, University of Electronic Science and Technology of China, Chengdu, Sichuan, 610054 China; he is also with Institute of Electronic and Information Engineering of UESTC in Guangdong, Guangdong, 523808, China (e-mail: yi.ding@uestc.edu.cn).
\IEEEcompsocthanksitem Dajiang Chen, Linpeng Gong, Mingsheng Cao, and Zhiguang Qin are with the Network and Data Security Key Laboratory of Sichuan Province, University of Electronic Science and Technology of China, Chengdu, Sichuan, 610054 China (e-mail: djchen@uestc.edu.cn; glpglp@std.uestc.edu.cn; cms@uestc.edu.cn; qinzg@uestc.edu.cn).
\IEEEcompsocthanksitem Guozheng Wu is with National Natural Science Foundation of China, Beijing, China (e-mail:wugz@nsfc.gov.cn).
\IEEEcompsocthanksitem Ning Zhang is with the Department of Computing Science, Texas A$\&$M University-Corpus Christi, Corpus Christi, TX 78412, USA (e-mail: ning.zhang@tamucc.edu).
}
}

\maketitle

\begin{abstract}
Internet of Medical Things (IoMT) can connect many medical imaging equipments to the medical information network to facilitate the process of diagnosing and treating for doctors. As medical image contains sensitive information, it is of importance yet very challenging to safeguard the privacy or security of the patient. In this work, a deep learning based encryption and decryption network (DeepEDN) is proposed to fulfill the process of encrypting and decrypting the medical image. Specifically, in DeepEDN, the Cycle-Generative Adversarial Network (Cycle-GAN) is employed as the main learning network to transfer the medical image from its original domain into the target domain. Target domain is regarded as a ``Hidden Factors" to guide the learning model for realizing the encryption. The encrypted image is restored to the original (plaintext) image through a reconstruction network to achieve an image decryption. In order to facilitate the data mining directly from the privacy-protected environment, a region of interest(ROI)-mining-network is proposed to extract the interested object from the encrypted image. The proposed DeepEDN is evaluated on the chest X-ray dataset. Extensive experimental results and security analysis show that the proposed method can achieve a high level of security with a good performance in efficiency.
\end{abstract}

\begin{IEEEkeywords}
Image encryption, Deep learning, Medical image, Internet of Medical Things
\end{IEEEkeywords}

\section{Introduction}
The Internet of Medical Things (IoMT) is an interdisciplinary filed which adopts the Internet of Things (IoT) technologies in the domain of medicine \cite{Gatouillat2018,Zhang2018,Chen2017File}. With the development of IoMT, many medical imaging equipments are widely connected to facilitate the process of diagnosing and treating for doctors, e.g., the brain magnetic resonance imaging (MRI) for brain tumor diagnosis and the computed tomography (CT)  of lung for lung nodule detection. In IoMT, medical images are usually managed by a system called Picture Archiving and Communication Systems (PACS) \cite{Liu2020}.
When a patient is scanned by the medical imaging equipment, the generated medical images will be firstly stored into the PACS. When the doctor begins to examine the patient, the PACS will retrieve the needed images from the database and transfer the images to the doctors' workstation which works with the patient information from the Hospital Information System (HIS). Although the PACS and HIS operate in an intranet environment, there are still some critical security issues when storing, transferring, and reviewing medical images, which preserve sensitive privacy information of patients. If an attacker, either an internal or an external attacker, has the ability to intrude the PACS or HIS, it becomes much easy to eavesdrop these medical images, resulting in severe privacy information leak of patients \cite{kzhang2018,Chen2018Channel,DChen2018}.

\begin{figure*}[t]
\centerline{\includegraphics[width=0.93\textwidth]{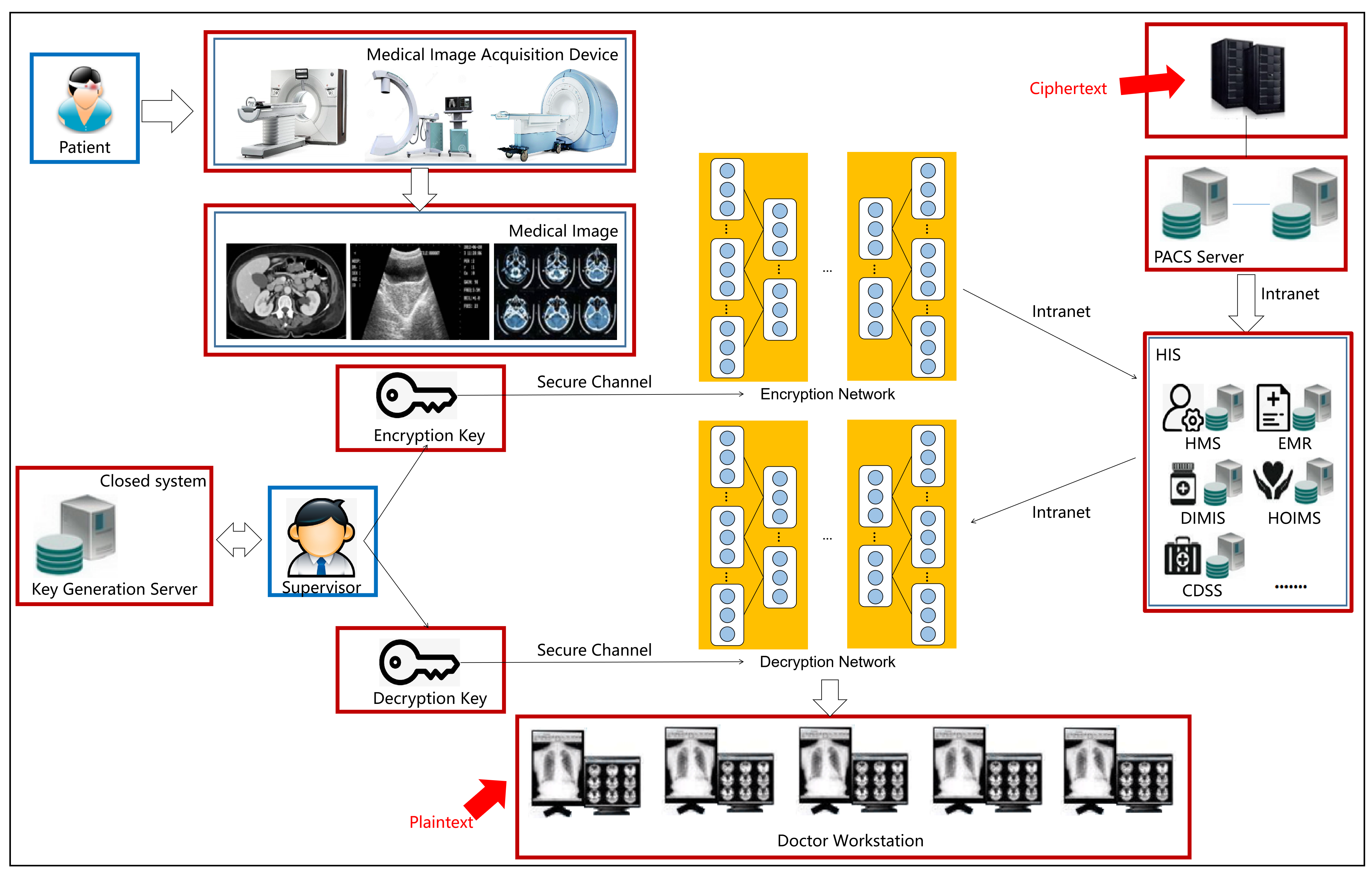}}
\caption{The architecture of DeepEDN.}
\label{fig:DLEDNet}
\end{figure*}

To safeguard the IoMT system and protect the patients' privacy, encryption and decryption can be performed on medical images, e.g., Data Encryption Standard (DES), Advanced Encryption Standard (AES) and the Hash Function~\cite{b1,b2}. In addition, image encryption based on chaotic systems are also employed in the literature~\cite{b3}. However, these methods are hard to achieve a good balance between the security performance and the encryption efficiency. Deep learning also holds great potential in dealing with this issue, where multi-layer neural networks extract a hierarchy of features from raw input images. The Convolutional Neural Network (CNN)~\cite{b4,Ale2019} has demonstrated the significant advantages in computer vision~\cite{b5,b6,b7,b8,add1,add2,add3,add4} as well as in image domain transfer~\cite{b9,b10}. Transferring the image from one  domain onto another can be considered as a problem of texture transfer where the goal is to learn the mapping relationship between an input image and an output image from a set of aligned image pairs. One of the most popular image-to-image transformation method is the Cycle-Consistent Adversarial Networks~\cite{b11}, which introduces two cycle consistency losses that transform the image from one domain to the other, and then reconstruct back to the original image. In fact, the deep learning algorithm has also been adopted to solve the problem of image denoising~\cite{b12}. 

Inspired by the above works, in this work, a deep learning based image encryption and decryption network (DeepEDN) is proposed for image-to-image transformation and image denoising. The novel idea is based on the following two important insights: (1) If the medical image can be transferred into other image domain which is greatly different from the original one, this medical image can be regarded as encrypted; and (2) the medical image decryption process can be implemented in the manner of image denoising or image reconstruction. In DeepEDN, the Cycle-GAN network is employed as the main learning network to implement the image-to-image transformation.
There are two domains in the encryption process: the original medical image domain and the target domain, where the target domain is regarded as a ``Hidden Factors'' to guide the learning model to realize the encryption process. For the encryption network, it consists of a generation network and a discriminator network. The former will generate the image similar to the target domain, while the latter will promote the generation network to generate the same images as the target domain by identifying the generated images. Therefore, after processing using the encryption network, the original medical image can be converted into the target domain and becomes the ciphertext. The decryption process is similar to traditional encryption-decryption methods, which is the inverse operation of the encryption process. In practice, a reconstruction network, which is actually a decryption procedure, is used to restore the encrypted image to the original one. In DeepEDN, the parameters of generation network is regarded as the private key for encryption while the parameters of reconstruction network is regarded as the private key for decryption. Moreover, DeepEDN adopts the unsupervised learning to train the learning network and it doesn't need much labeled samples. It overcomes the dataset issues in training and is beneficial to the application of deep learning in cryptography filed.


Based on DeepEDN, the PACS system is improved by employing a key generation server. As shown in Fig.\ref{fig:DLEDNet}, the key generation server is in charge of training the encryption network and the decryption network. The PACS system can call the encryption network to encrypt the medical image and then store these ciphertext images into the image database. When reviewing, the HIS system will adopts the decryption network to decrypt the ciphertext image to the original one. The encryption network and the decryption network will be transferred over the secure channel. Moreover, a ROI-mining-network is proposed to directly extract the ROI (organ or tissue) from the encrypted medical image without decryption. More specific, when inputting an encrypted medical image into the ROI-mining-network, the interested segmented object can be directly extracted without revealing other parts of the patient's information.

In a nutshell, the main contributions of this work are summarized as follows:

1. An novel medical image encryption and decryption network, DeepEDN, is developed to realize the encipherment process by applying the deep learning in the field of image-to-image transformation. The proposed encryption method is with the large key space, one-time pad, and be sensitive to key change. To the best of our knowledge, this work is the first work to attempt to adopt the deep learning method in the area of medical image encryption.

2. A ROI-mining-network is proposed to directly extract the interested segmentation region from the encrypted medical image instead of decrypting the ciphertext image firstly. From to the experiments, it can be found that the proposed approach can realize the data mining process directly from the privacy-protected environment.

3. Extensive experiments are conducted on the chest Xray dataset  to evaluate the proposed DeepEDN. The results demonstrate that the medical image can be transmitted  with a high level of security and efficiency, compared with existing medical image encryption methods. Moreover, the proposed encryption algorithm can resist various attacks, even if the attacker has known the complete process for key generation.

The remainder of this paper is as follows. Section~\ref{sec:related work} gives an introduction of the image encryption and deep learning. Section~\ref{sec:endecryptnet} presents the details of the proposed DeepEDN. Section~\ref{sec:securityAnalysis} analyzes the security of proposed method. Section~\ref{sec:experiment} shows the encryption and decryption performance and evaluates the network efficiency. Section~\ref{sec:conclusion} gives an summarization.


\section{Related Work} \label{sec:related work}
\subsection{Medical Image Encryption}
In the literature, there are many approaches proposed for image encryption~\cite{b14,b15,b16}.
Based on the transformation of cosine number, Lima \emph{et al}~\cite{b14} proposed a novel scheme for encrypting the medical images, in which modular arithmetic was required to prevent rounding-off errors.
It is a flexible method which can be applied to the medical images in the format of DICOM.
 Natsheh \emph{et al} \cite{b15} proposed a simple yet effective encryption approach for multi-frame DICOM medical images, which leveraged AES to accelerate the encryption and decryption process.
In \cite{b16}, Mukhedkar \emph{et al} demonstrated that image encryption can be done by using a faster Blowfish Algorithm and image hiding technique.

Medical image encryption can also be performed using chaotic maps \cite{b17,b18,b19}.
In \cite{b17}, a novel selective chaos-based image encryption scheme was proposed to encrypt the medical image.
Chong \emph{et al}~\cite{b18} presented a novel chaos-based medical image encryption scheme, in which, a bit-level shuffling algorithm was used as a substitution mechanism in the process of permutation.
In \cite{b19}, an image encryption algorithm was proposed to effectively protect the image, which is based on wavelet function and four-dimension chaotic system.
The wavelet function was used to scramble the pixel location in the image, while the four-dimension chaotic system was used to disturb the pixel value.

\subsection{Image-to-Image Transfer by Generative Adversarial Networks}
Based on the creative work of GAN network proposed in \cite{b20}, many researchers put their focus on the usage of GAN-based methods in different applications.
The GAN network is consist of a generator and an adversarial discriminator.
The former one is used to capture the data distribution, while the latter one evolves to distinguish the fake data from the real data~\cite{b20}.
GAN-based methods can achieve state-of-the-art results in wide variety of applications such as image generation~\cite{b21}, image segmentation \cite{b22}, image super-resolution ~\cite{b23}, and image-to-image transformation ~\cite{b24}\cite{b25}.

In \cite{b27}, a conditional generative adversarial network (CGAN) was  used  to implement the image-to-image transformation.
This approach achieved a better performance on synthesizing photos from label maps, reconstructing objects from edge maps and colorizing images.
In \cite{b26}, DualGAN mechanism was proposed to train the learning network from two sets of unlabeled images.
Taking two sets of unlabeled images as the input, DualGAN learned two reliable image transformation networks at the same time and hence can facilitate a lot of image-to-image transformation tasks.
In~\cite{b11}, Cycle-GAN was proposed to fulfill image transformation task with unpaired images.
The Cycle-GAN simultaneously trained two sets of GAN models: one model was used to learn mapping from class A to class B
 and the other one learned the mapping from class B to class A.
The loss was redefined with the combination of these two mappings.
The success of GAN was depend on the idea of an adversarial loss, which forces the generated images to distinguish from target images.

Regardless of their merits, these algorithms are difficult to achieve a good balance between the security and the efficiency.
On the one hand, as there is plenty of information in a single medical image with high correlation among these information, when encrypting the medical image, the block encryption algorithm is with low efficiency and cannot meet the real-time requirement.
On the other hand, the chaotic system usually adopts the one-dimensional chaotic map to generate pseudorandom sequences.
Consequently, the chaotic system tends to be easy to analyze and predict through a nonlinear prediction method based on phase-space reconstruction\cite{b27}.
The deep learning algorithm has been employed in the security field~\cite{b28,Chen2018An,b29}.
However, there is no work on the medical image encryption and decryption.

In this paper, deep learning techniques are used to encrypt and decrypt medical images,
 in which parameters of the deep learning network model are regarded as the encryption and decryption keys.
Due to the large key space and the complex model structure, the proposed method can achieve a high level of security with a high efficiency.
\section{Encryption and Decryption Network} \label{sec:endecryptnet}

\begin{figure}[t]
\centerline{\includegraphics[width=0.48\textwidth]{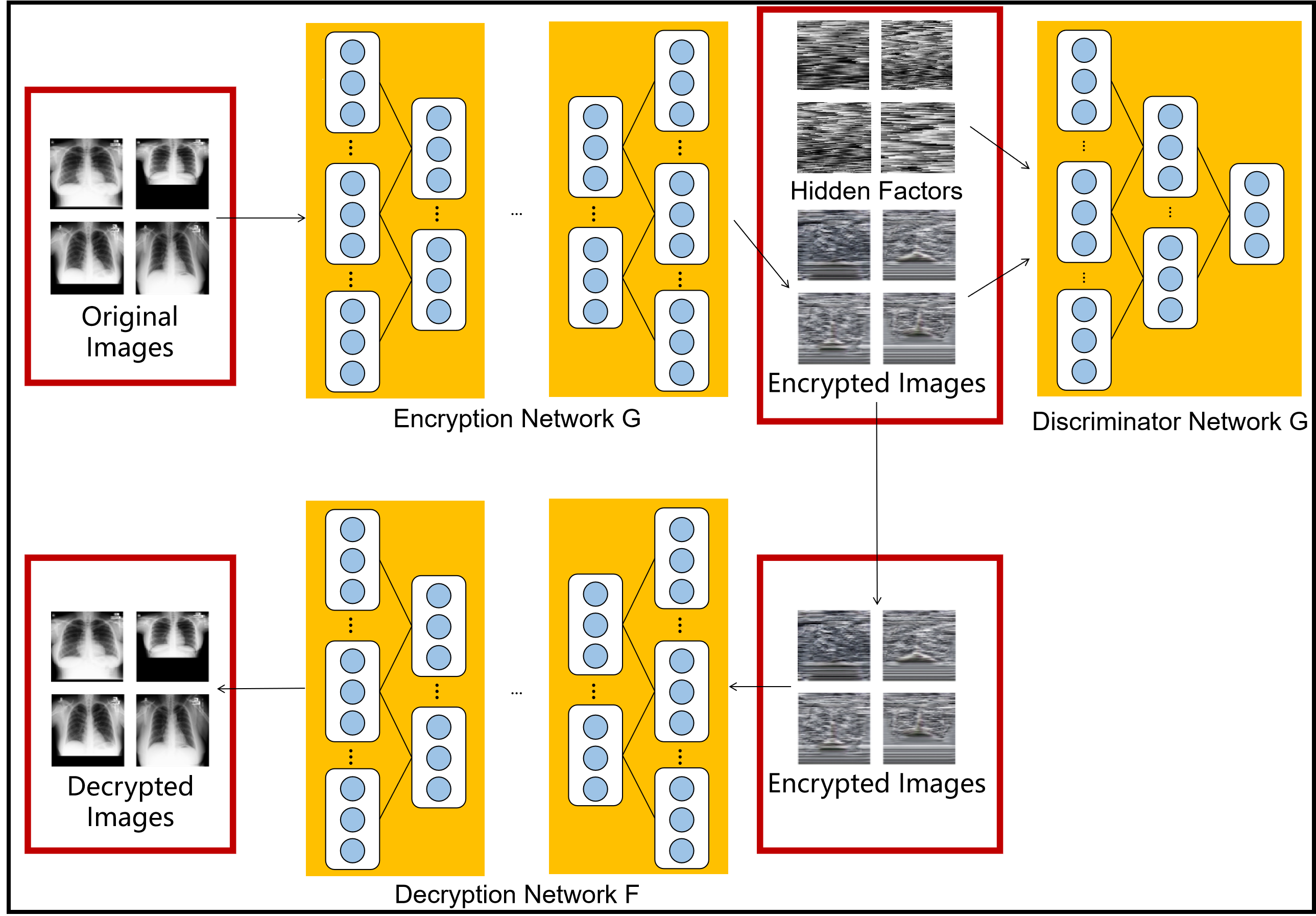}}
\caption{The overall framework of DeepEDN.}
\label{fig:overallDLEDNet}
\end{figure}

\subsection{Architecture of DeepEDN}

As shown in Fig.\ref{fig:overallDLEDNet}, DeepEDN mainly consists of three subnetworks: the encryption network G, the discriminator network D, and the decryption network F. The encryption network G is used to encrypt the original input images, the decryption network F is responsible for restoring the encrypted images to the original one (decrypting the image), and the discriminator network D is mainly designed for improving the performance of the encryption network by distinguishing the generated images from the images in target domain (Hiding Factors). In deep learning methods, loss function is usually used to train the model. The overall loss L of the proposed model is given as follows:
\begin{equation}
L = L_G+L_D+L_R\label{eq},
\end{equation}
where the $L_G$ indicates the loss of the encryption network G, $L_D$ indicates the loss of the discriminator network D, and $L_R$ indicates the loss of the decryption network F.
\subsubsection{Encryption Network and Decryption Network}
The encryption network G is used to transform the original medical images into the target domain for medical image encryption. The G network begins with an initial convolution stage to spatially downsampling and encode the images, and the useful features obtained in this stage will be used for the following transformation. Then, nine residual blocks[48] are performed to construct the manifold and content features. The output images are reconstructed with two up-convolution blocks which contain a strided convolutional layer and the stride is set to 2. Finally, the prediction is exported by a 7$\times$7 convolution kernel. In addition, the structure of decryption network F is the same as the encryption network G.

The proposed model includes two mappings $G : X\to Y$ and $F: Y\to X$. The goal of mapping function G is to learn how to transform the original medical images $X$ into the images $Y$ in target domain, and cheat the discriminator network D. When the discriminator network D cannot successfully distinguish whether an image is generated by the encryption network G or a real ciphertext image domain $Y$, it means that the encryption network G converts the original patient image domain $X$ into a ciphertext image domain $Y$ successfully. The loss $L_G$ of the encrypted network G is:
\begin{equation}
L_G = min_{G}(E_{x \sim pdata(x)}{log(1-D(G(x))})\label{eq},
\end{equation}
where G represents an encryption network, and D represents the discriminator network. The goal of $L_G$ is to minimize the success rate of the discriminator network D for detecting the ciphertext generated by the encryption network G. In addition to the encryption, another goal of the proposed method is to ensure that the restored image reserves the texture information of the original one even it is encrypted. As shown in Fig.\ref{fig:overallDLEDNet}, for each image $x$ from domain $X$, the reconstruction loss measures the difference between $G(x)$ and the original image, i.e., $x\rightarrow G(x)\rightarrow F(G(x))\approx x$. The reconstruction loss L is defined as:
\begin{equation}
	\begin{split}
	L_R &= E_{x \sim p_{data(x)}}{||F(G(X) - X||_1}\\
	&= E_{x \sim p_{data(x)}}\sum_{i=1}^n|F(G(x_i) - x_i|
    \end{split}
\end{equation}

\subsubsection{Discriminator Network}

The discriminator network D is used to evaluate whether the output image of encryption network belongs to the target domain.
For the discriminator network D, after processing with initial convolutional layers, two strided convolutional blocks are adopted to reduce the resolution of image and to encode essential local features for subsequential discrimination. Then, the network employs a feature construction block and a 3$\times$3 convolutional layer to obtain the final result. In addition, for each convolutional layer, the Leaky ReLU (LReLU) with $\alpha=0.2$ is adopted and followed with a batch normalization (BN) layer.

The training of the discriminator network D is to classify the images and check whether it comes from the ciphertext domain $Y$
 or is generated by the encryption network G. The encryption network G attempts to generate an image $G(x)$ that is similar with the image in domain $Y$, while the discriminator network D aims to find the difference between transformed samples from $G(x)$ and real samples in $Y$.
Minimizing loss $L_D$ of the discriminator network D is equivalent to maximizing the classification accuracy of the discriminator network D, which is opposite to the goal of the encryption network G. The loss $L_D$ given as follows:
\begin{equation}
{L_D = E_{x\sim pdata(x)}{log{D(x)}}+E_{x\sim pdata(x)}{log(1-D(G(x)))}}, \label{eq}
\end{equation}
where G represents the encrypted network, and D represents the discriminator network. $L_D$ and $L_G$ in the GAN network form a adversarial relationship. When the two networks reach an equilibrium state, the discriminator network D can achieve $50\%$ classification accuracy for both the generated ciphertext image and the real ciphertext domain image $Y$. In other words, the ciphertext image generated by the encryption network G is very similar to the real ciphertext domain $Y$ so that the discriminator network D cannot distinguish them.

\subsubsection{The key generation process}
In DeepEDN, the final parameters of network G can be considered as the private key for encryption while the parameters of network F are regarded as the private key for decryption.
\begin{figure*}[t]
\centerline{\includegraphics[width=0.93\textwidth]{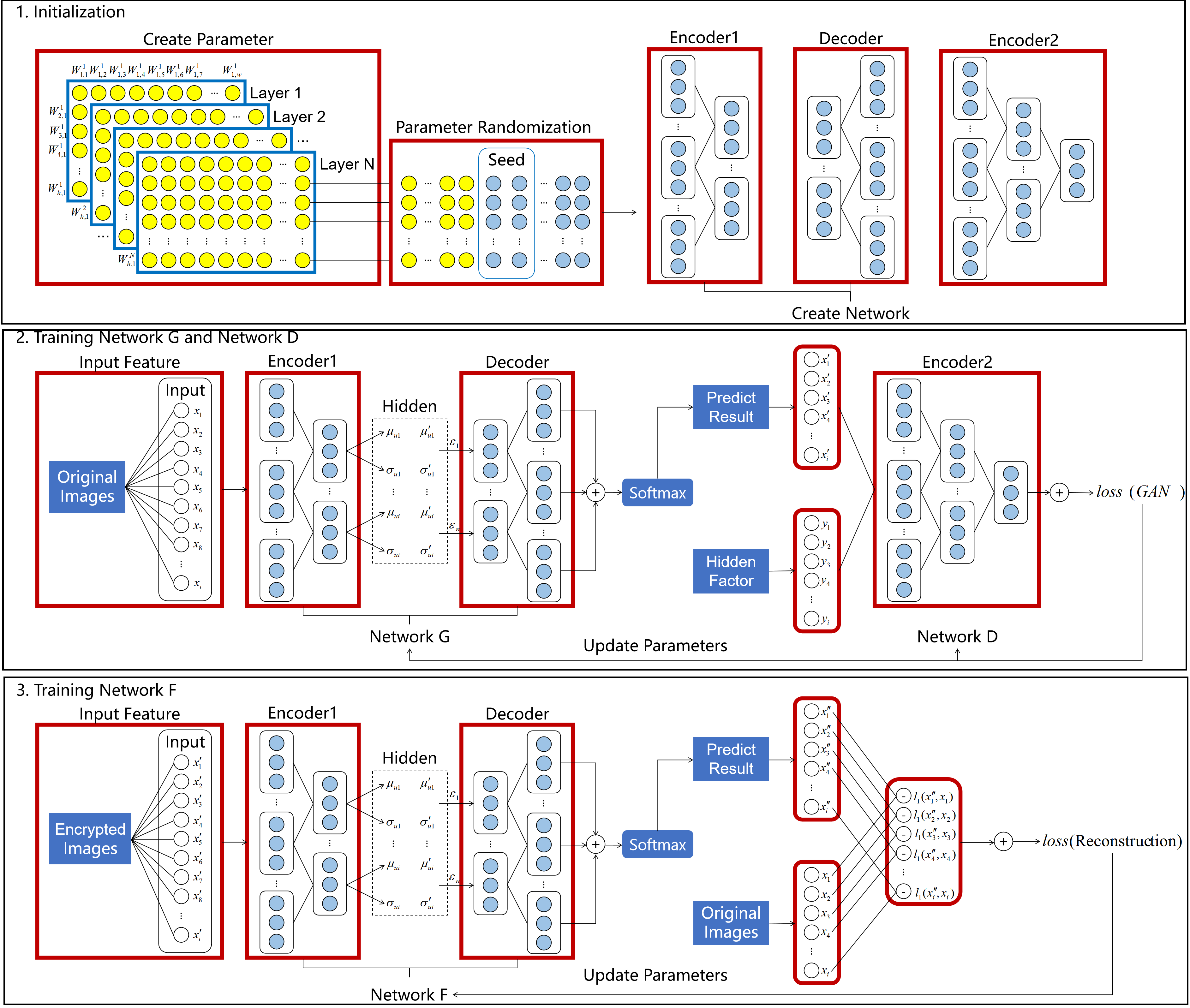}}
\caption{The key generation process.}
\label{fig:keygeneration}
\end{figure*}
For encryption, the parameters for each convolutional layer are firstly randomly initialized as follows:
\begin{equation}
W_{n} = random[w_{n, 1}, w_{n, 2},...,w_{n, j},...]\label{eq},
\end{equation}
where $w_n$ is the $n^{th}$ convolutional layer and $w_{n,j}$ is the $j$-$th$ parameter of one convolutional layer. Therefore, the private key $W$ for encryption is actually composed of all the parameters of each convolutional layer, and is defined as follows:
\begin{equation}
W = consist[W_1, W_2,...,W_n,...]\label{eq}
\end{equation}
When training the encryption network, the private key for encryption is continuously updated and refined with different input images through forward  propagation training process. The adversarial loss $L_{gan}$ is calculated to measure the difference between the predicted result and the target one in ``Hidden Factors'', thereby guiding the network to train and update the private key for encryption.

Except for the forward propagation, the back-propagation algorithm (BP) is also employed to pass loss of the entire network back to the convolutional layers. It is actually a gradient descent, which can further update the parameters in each layer to achieve better performance. The gradient descent can be described as:
\begin{equation}
	\begin{split}
\theta_{j}=&\theta_{j} - \alpha \lor J(\theta)= \theta_{j} - \alpha\frac{\delta}{\theta_{j}}J(\theta)\\
    = &\theta_{j} - \alpha\frac{\delta}{\theta_{j}}\frac{1}{2m}\sum_{i=1}^m(h_{\theta}(x^{i})-y^{i})^2\\
    =& \theta_{j} - \alpha\frac{1}{2m}\sum_{i=1}^m2\frac{\delta}{\theta_{j}}(h_{\theta}(x^{i})-y^{i})(\frac{\delta}{\theta_{j}}(h_{\theta}(x^{i})-y^{i}))\\
    =&\theta_{j} - \alpha\frac{1}{m}\sum_{i=1}^m(h_{\theta}(x^{i})-y^{i})(\sum_{i=0}^n\frac{\delta}{\theta_{i}}\theta_{i}x_i-\frac{\delta}{\theta_{i}}y^{i})
	\end{split}
\end{equation}
where $\theta_{j}$ is the value of parameter $\theta$ in the $j^{th}$ training epoch. $\alpha$ is the learning rate and $\lor J(\theta)$ means the gradient that is passed back to the convolution layer $\theta$ in the $j^{th}$ training epoch.

The generation process of the private key for decryption is similar to the process of generating the privacy key for encryption, except that the initial input of the decryption network is the predicted result of the encryption network. In addition, the loss of decryption network is the reconstruction loss, which is given in Equ. (8).
\begin{equation}
	\begin{split}
	L_{reconstruction}
=& E_{x \sim p_{data(x)}}{||F(P(X))-O(X)||_1}\\
=& E_{x \sim p_{data(x)}}\sum_{i=1}^n|F(P(x_i)) - O(x_i)|
    \end{split}
\end{equation}
where $F()$ is the decryption network, $P(x)$ is the pixel $x$ in the predicted image, and $O(x)$ is the corresponding position pixel $x$ in the original image. The encryption network G and the decryption network F are trained in an alternative manner. When the loss becomes stable, the final parameters (privacy keys) for encryption and decryption network can be obtained.
The complete privacy key generation process is presented in Fig.\ref{fig:keygeneration}.

After obtaining the key, the patient's medical image can be encrypted by the encryption network G, and then decrypted by the decryption network F.
The proposed medical image encryption/decryption algorithm is given in Alg. \ref{code:recentStart}.

\renewcommand{\algorithmicrequire}{\textbf{Initialization:}} 
\renewcommand{\algorithmicensure}{\textbf{Output:}} 
\begin{algorithm}[ht]
\caption{Image Encryption/Decryption.}
\label{alg:Framwork}
\begin{algorithmic}[1]
\small{
\Require
Digitize the 255$\times$255 image into a 255$\times$255 matrix X$_*^{0}$.And then enter it into our 21-layer($L_c$) encryption/decryption model.
\While {L$<L_c$}
\ForAll {element (X$_1^L$, X$_2^L$,...) in matrix X$^L$}
\State Each pre-trained 3$\times$3 convolution kernel W$_*^L$ in L$^{th}$ layer sequentially traverses the image matrix and multiplies it with the corresponding elements of the matrix(W$_*^L\times$X$_*^L$).
\State Add the obtained nine W$_*$X$_*$ to get a new predicted value X$_*^{L+1}$ in (L+1)$^{th}$.
\State Collect all X$^{L+1}$ and combine them into a new matrix to form the next-level feature matrix.
\EndFor;
\State $L=L+1$;
\EndWhile
\Ensure
Convert the last layer of matrix X$^{L_c}$ into an image to get the final encrypted/decrypted image.
}
\label{code:recentStart}
\end{algorithmic}
\end{algorithm}
Since the GAN model is highly nonlinear and randomly initialized, and the parameters of the learning network can be totally different at different training times.
In other words, the GAN network is unstable, which is its weakness when used for computer vision tasks.
However, this instability has advantages for cryptography. By utilizing this instability, the proposed deep learning based encryption method can be regarded as an one-Time Pad (OTP) method.
Specifically, the parameters of encryption network are totally different after training the network at different times.
Overall, due to the depth and complex structure of the learning encryption network, the proposed framework is with higher security.


\subsection{ROI Mining Network in Ciphertext Environments}
Although various methods have achieved a good performance in protecting the image privacy, it is still a challenge to directly obtain the effective information in a ciphertext environment, e.g., extracting the desired ROI from the encrypted medical image. In DeepEDN, a ROI-mining-network is proposed to segment the region of interest from the encrypted medical image. In order to extract useful texture features in a ciphertext environment, a deeper network structure is adopted to learn semantic features to accurately segment the specific target. The input encrypted image will be processed with 5 blocks, and each block has a down-sampling convolution. In the first block, since the convolutional kernel size is set to 3$\times$3, each convolution operation can learn the local information from the input image. As the increasement of the network depth, more abstracted semantic information can be obtained. Finally, by combining the output results from each convolution layer, the final prediction results can be achieved.

Each block in ResNet-50 has two sub-blocks. One is the identity (ID) block in which the stride of each convolution layer is 1. The identity block is mainly used to extract abstract features through multi-layer convolution. Since the dimensions of the input and output are the same, these feature maps can be serially connected. The other basic block is Conv Block where the dimensions of input and output are different and it is used to change the dimension of the feature vector and to resize feature size through a strided convolutional layer.
The CNN-based neural network commonly converts the image into a small feature map with many channels.
However, by increasing the network layers, there will be a huge number of output channels and parameters, resulting in increased computational complexity and reduced network efficiency. Therefore, it is necessary to reduce the dimension of Conv Block before processing with the Identity Block.

In DeepEDN, the proposed ROI-mining-network is used to implement the medical image segmentation task in the ciphertext environment.
The medical image segmentation is a key step in medical image analysis.
Its purpose is to extract useful features and segment the doctors' interested objects.
The segmentation results can provide a reliable basis for clinical diagnosis and pathological research. When training the ROI-mining-network, the encrypted medical image is firstly used as the input of the network. Then, the pixel-level segmentation labels in the corresponding medical image are adopted to supervise the training process. Finally, the model parameters are updated by the mean square error (MSE). The loss function of this segmentation model is described as:
\begin{equation}
L_{S} = \frac{1}{N}\sum_{i=0}^N (g_i - p_i)^2\label{eq}
\end{equation}
where $g_i$ represents the value of the $i^{th}$ pixel in the label and $p_i$ is the predicted value of the $i^{th}$ pixel in the predicted result. $N$ represents the total number of pixels in this image. The final training result is a high-quality splitter that can segment the medical images without decryption.

The usage of the ROI-mining-network is of great significance for medical image security.
It can implement data mining in an untrusted environment to securely extract specific objects, which is also beneficial for protecting the privacy of patient. This network can further improve the security of  medical image analysis and can be widely used in many medical applications.

\subsection{Adversary Model}

In DeepEDN, the most important factors of key generation process include the structure of the model and the chosen hidden factors.
If the network structure or hidden factors leaks, the attacker can train a similar encryption network by imitating the private key generation process so as to crack the ciphertext image.
This kind of attack is called as the imitation learning attack. This paper proposes three possible adversary models for imitation learning attack: the hidden factors leakage, the network architecture leakage, and both the hidden factors and the network architecture leakage.

\subsubsection{Hidden Factors Leakage}
Hidden factors leakage means that the attacker has known the hidden factors used for the encryption, and tries to employ  the same hidden factors to train the attacking network with several different network architectures to decrypt the ciphertext image.
There are two encryption and decryption networks with different network structures: the encryption/decryption network A and encryption/decryption network B.
These two encryption and decryption networks are trained with the same hiding factor.
If the decryption network B is able to recover the image encrypted by the encryption network A, it means that the attacker can crack the secure key by imitation learning attack.

\subsubsection{Network Architecture Leakage}
Network architecture leakage assumes that only the architecture of the encryption and decryption network is leaked, and the hidden factors remains confidential.
In this adversary model, the attacker can decrypt the encrypted image by training the same network structure without knowing the hidden factors.
The attacker can employ different hidden factors to train the same network structure to construct different decryption networks. If the attacker is able to recover the encrypted ciphertext image, the attack is successful.


\subsubsection{Both Hidden Factors and Network Architecture Leakage}
The strongest adversary model is that both the network architecture and hidden factors are leaked.
In such a scenario, the attacker can train the network with the same network structure and hidden factor adopted for training the encryption/decryption network.
To prevent such attacks, after each training of the network, the parameters of the encryption/decryption network representing the actual private key must be completely different. It means that the proposed encryption algorithm should be similar to the OTP and can be regarded as a chaotic encryption algorithm.

\begin{table*}[htbp]
\caption{The structure of encrypted network and decryption network.}
\begin{center}
\begin{tabular}{|c|c|c|c|c|c|c|}
\hline
\textbf{Convolution Layer Name}& \textbf{Number}& \textbf{Size}& \textbf{Input Channels}& \textbf{Output Channels}& \textbf{Parameters}& \textbf{Total Parameters}\\
\hline
Down Convolution1&1&7$\times$7&3&32&4704&4704\\
\hline
Down Convolution2&1&3$\times$3&32&64&18432&23136\\
\hline
Down Convolution3&1&3$\times$3&64&128&73728&95864\\
\hline
Residual Blocks&18&3$\times$3&128&128&2564208&2661072\\
\hline
Up Convolution1&1&3$\times$3&128&64&73728&2734800\\
\hline
Up Convolution2&1&3$\times$3&64&32&18432&2753232\\
\hline
Up Convolution3&1&7$\times$7&32&3&4704&2757936\\
\hline
\end{tabular}
\label{tab01}
\end{center}
\end{table*}

\begin{table*}[htbp]
\caption{The structure of ROI-Mining-Network.}
\begin{center}
\begin{tabular}{|c|c|c|c|c|c|c|}
\hline
\textbf{Convolution Layer Name}& \textbf{Number}& \textbf{Size}& \textbf{Input Channels}& \textbf{Output Channels}& \textbf{Parameters}& \textbf{Total Parameters}\\
\hline
Block 1&2&7$\times$7&3&64&4704&4704\\
\hline
Block 2&3&3$\times$3&64&256&18432&23136\\
\hline
Block 3&12&3$\times$3&256&512&73728&95864\\
\hline
Block 4&18&3$\times$3&512&1024&2564208&2661072\\
\hline
Block 5&1&3$\times$3&1024&2048&73728&2734800\\
\hline
\end{tabular}
\label{tab02}
\end{center}
\end{table*}

\section{Security Analysis} \label{sec:securityAnalysis}
In DeepEDN, both encryption and decryption network are constructed with 24 layers and the number of the parameters for each network is 2,757,936.
The explicit specification of the network are represented in Table \ref{tab01}.
For the ROI-mining-network, a deeper resnet-50 architecture is adopted.
The network structure of the ROI-mining-network is given in Table \ref{tab02}.
The dataset is the chest x-rays\cite{b31}. The proposed method is running on the Nvidia GTX 2080Ti.
When training the network,  it takes around 10 mins for each epoch of the model.

\subsection{Key Security Analysis}
The ideal encryption scheme has the following characteristics: 1) the key space is large enough so that it can effectively resist the exhaustive attack under the premise of the existing computing power; 2) the key generated for each time should be different, i.e., the key generation should be uniform at random; and 3) the encrypted image must be highly sensitive to the key. The security of the key will be analyzed from these three characteristics in the following sections.
\subsubsection{Key Space Analysis}
The size of the key space determines the difficulty of an attacker encounters using an exhaustive attack. In this work, the key space of the proposed encryption algorithm is the number of parameters for the deep learning network, with a total of 2,757,936 parameters in the experiments.
Each parameter or key is a floating point number between 0 and 1, which is 32 bits in the computer and can be expressed as a decimal number with 10 significant digits.
Therefore, the key space of the encryption model can be expressed as the  $(2^{32})^{2757936}$. It becomes very hard for attackers to break system and the proposed scheme can effectively resist attacks.

\subsubsection{Key Randomness Analysis}
\begin{figure}[t]
\centerline{\includegraphics[width=0.48\textwidth]{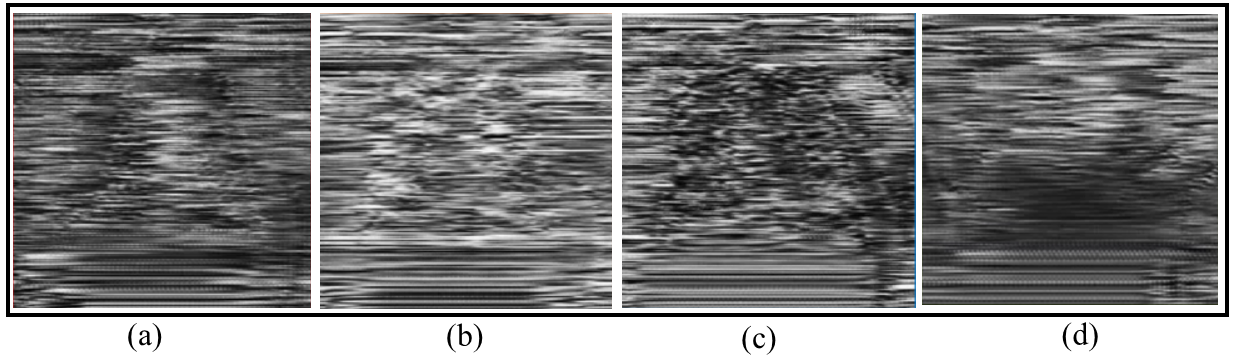}}
\caption{The same image is encrypted with the key obtained from four networks.}
\label{fig:encryptedimage4}
\end{figure}
The encryption network is trained four times with the same settings.
Accordingly, the parameters of these four networks are adopted as encryption keys, i.e., Key A, Key B, Key C and Key D, respectively.
The same image is encrypted with these four keys, and the encrypted images are shown in Fig. 11.
Fig. \ref{fig:encryptedimage4}(a), Fig. \ref{fig:encryptedimage4}(b), Fig. \ref{fig:encryptedimage4}(c), and Fig. \ref{fig:encryptedimage4}(d) are the results obtained by encrypting the same original image from four networks.
It is clear that these four images are different. The similarity among these four encrypted images (SSIM) are calculated, and the result can be found in Table \ref{tab03}.
The SSIM index between different images is mostly lower than 0.1 which indicates that the similarity between different images is very low.

According to the experiment, it can be found that since parameters of the neural network are randomly initialized, the private keys for the medical image encryption network are totally different after every training. These difference results in different encrypted images which are processed with different encryption networks. The idea behind this is the training of the deep learning network is not stable. Different initialized parameters can lead to dramatic difference in final parameters in different training. It can be demonstrated that the proposed method is similar to OTP and can be regarded as one type of OTP method.
\begin{table}[htbp]
\caption{SSIM between two encrypted images.}
\begin{center}
\begin{tabular}{|c|c|c|c|c|}
\hline
\textbf{Image} &\textbf{A} &\textbf{B} &\textbf{C} &\textbf{D} \\
\hline
a &1 &0.07 &0.11 &0.09 \\
\hline
b &0.07 &1 &0.08 &0.04 \\
\hline
c &0.11 &0.08 &1 &0.05 \\
\hline
d &0.09 &0.04 &0.05 &1 \\
\hline
\end{tabular}
\label{tab03}
\end{center}
\end{table}

\subsubsection{Key Sensitivity Analysis}
Unlike traditional encryption algorithms, the error in deep learning models will be propagated among layers. In the convolution process, the $l^{th}$ pixel in the $N^{th}$ layer feature map is passed to a neighboring pixel of the $(N+1)^{th}$ layer via a $3\times3$ convolution kernel. When a feature point is erroneous, it will be passed to the $3\times3$ feature points in the next layer. As the depth of the convolutional network increases, the error of feature points will increase with two pixels for each laye. In the up-sampling process, this error increases exponentially with the superposition of the deconvolution operation. The experiment assumes the attacker knows the most private keys. And only about $5\%$ of key parameters are modified which is regarded as the unknown part. Then, the encrypted image is input to the network with new parameters, and the network cannot decrypt the ciphertext image to the original one. It means that even if only $5\%$ of the parameters is changed, the private key cannot encrypt or decrypt the medical image correctly. In other words, it becomes very hard for attackers to guess at least $95\%$ of the right key parameters in a key space with $(10^{10})^{2757936}$ so as to break the proposed algorithm.

\subsection{Ciphertext Security Analysis}

\subsubsection{Histogram Analysis}
\begin{figure}[t]
\centerline{\includegraphics[width=0.48\textwidth]{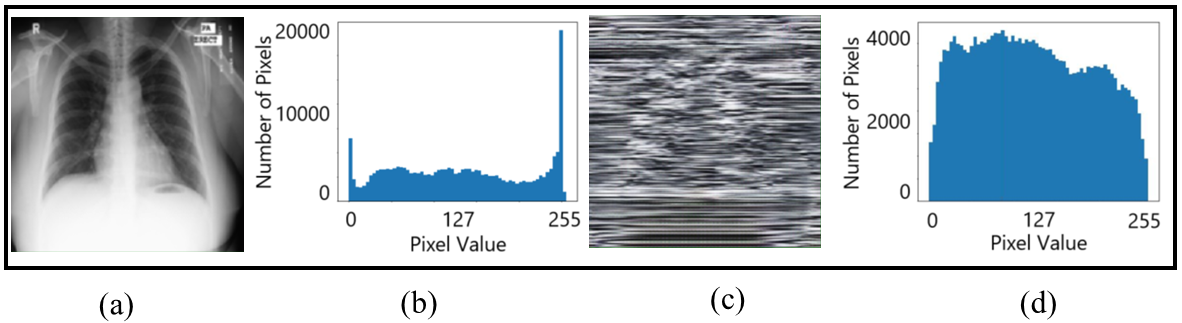}}
\caption{Pixel distribution of the original image and the encrypted image.}
\label{fig:pixeldistribution}
\end{figure}

To evaluate the performance of the proposed encryption network, The original image is shown in Fig. \ref{fig:pixeldistribution}(a) and the encrypted image is shown in Fig. \ref{fig:pixeldistribution}(c). Through the experiment, it can be found that the pixel distribution of the original image and the encrypted image is quite different. In Fig. \ref{fig:pixeldistribution}, the pixel histogram of the original chest X-ray image has a total of 57600*(240 * 240) pixels (Fig. \ref{fig:pixeldistribution}(b)), in which more than 30,000 pixels have a value of 0, and more than 5000 pixels have a value of 255. The pixel distribution of original image is relatively concentrated. However, the distribution of encrypted medical images (Fig. \ref{fig:pixeldistribution}(d)) is more uniform, which is helpful for mitigating the statistical analysis.

\subsubsection{Entropy Analysis}
The information entropy of the encrypted image is regarded as an effective quantitative measurement for algorithms to against statistical attacks. The image information entropy represents the statistical feature of the grayscale distribution of the image. In an ideal case, the encrypted image should be similar to random noise, the grayscale distribution tends to be uniform, and the expected value should be 8. The information entropy formula is defined as follows:
\begin{equation}
Entropy = -\sum_{l=0}^N p(l)log_2 (p(l))\label{eq}
\end{equation}
where $N$ is the number of gray levels of the pixel value and $p (l)$ is the probability that the pixel value $l$ appears.
The entropy metric is calculated on the encrypted medical image, and the results are given in Table \ref{tab04}.
It is clear that the image encrypted by the proposed method is close to the ideal value of 8 on information entropy. Experiments show that the images encrypted by the proposed method has the ability to resist the statistical attacks.
\begin{table}[htbp]
\caption{Evaluation of the Entropy Effect of Our Network.}
\begin{center}
\begin{tabular}{|c|c|c|c|c|c|}
\hline
\textbf{Image Id} &\textbf{1} &\textbf{2} &\textbf{3} &\textbf{4} &\textbf{5}  \\
\hline
Entropy &7.96 &7.96 &7.95 &7.94 &7.95  \\
\hline
\hline
\textbf{Image Id}  &\textbf{6} &\textbf{7} &\textbf{8} &\textbf{9} &\textbf{10} \\
\hline
Entropy &7.97 &7.95 &7.96 &7.96 &7.95 \\
\hline
\end{tabular}
\label{tab04}
\end{center}
\end{table}

\subsection{Security Analysis under Different Adversary Models}
The experiments are conducted to validate whether an attacker can generate a key under three different adversary models.
\subsubsection{`Hidden Factors Leakage}
In this experiment, four different network structures are considered, namely network A, network B, network C and network D.
The training conditions are kept the same. The network structure of these four networks are shown in Table \ref{tab05}.

\begin{table}[htbp]
\caption{The network model of the different architectures.}
\begin{center}
\begin{tabular}{|c|c|c|c|c|}
\hline
\textbf{Convolution Layer}& \textbf{Net. A}&\textbf{Net. B}&\textbf{Net. C}&\textbf{Net. D}\\
\hline
Down Convolution1&1&1&1&1\\
\hline
Down Convolution2&1&1&1&1\\
\hline
Down Convolution3&1&1&1&1\\
\hline
Residual Blocks&18&15&12&9\\
\hline
Up Convolution1&1&1&1&1\\
\hline
Up Convolution2&1&1&1&1\\
\hline
Up Convolution3&1&1&1&1\\
\hline
\end{tabular}
\label{tab05}
\end{center}
\end{table}

\begin{figure}[t]
\centerline{\includegraphics[width=0.48\textwidth]{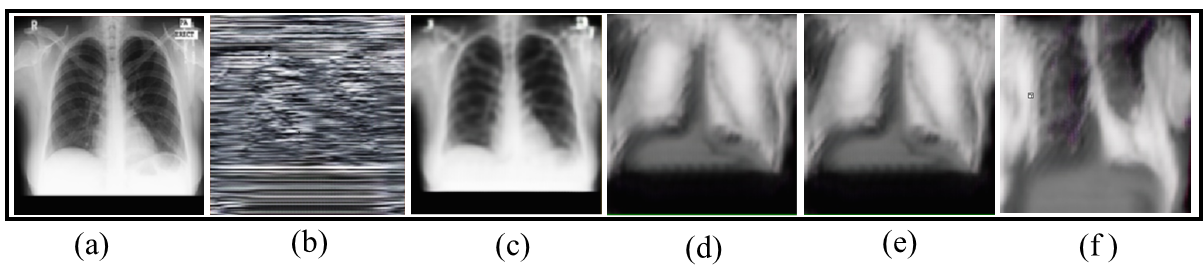}}
\caption{The decryption performance for different networks.}
\label{fig:performanceonnetwork}
\end{figure}

The original image is encrypted by using the trained network A. The ciphertext image is then decrypted by the decryption network obtained from network A, network B, network C and network D respectively to restore the original image. As shown in Fig.\ref{fig:performanceonnetwork}, the original image (Fig.\ref{fig:performanceonnetwork}(a)) encrypted by network A (the encrypted image is shown in Fig.\ref{fig:performanceonnetwork}(b)), can only be correctly decrypted by the decryption network A as shown in Fig.\ref{fig:performanceonnetwork}(c). While the image decrypted by network B, network C and network D are visually unrecognizable and the result is shown in Fig.\ref{fig:performanceonnetwork}(d), Fig.\ref{fig:performanceonnetwork}(e), Fig.\ref{fig:performanceonnetwork}(f), respectively. Experiments show that even if the attacker knows the hidden factors, the ``attack network'' trained with different network structure still cannot be used to decrypt the ciphertext image.

\subsubsection{Network Architecture Leakage}
\begin{figure}[t]
\centerline{\includegraphics[width=0.5\textwidth]{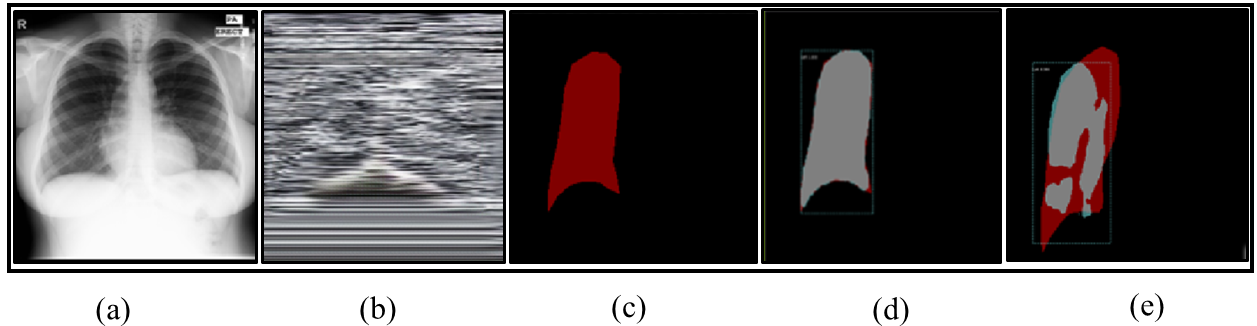}}
\caption{The mutual decryption performance between networks under different hidden factors training.}
\label{fig:mutualperformancenetwork}
\end{figure}

In this experiment, different hidden factors are adopted to train the encryption network with the same network structure. All training conditions are kept the same. As shown in Fig.\ref{fig:mutualperformancenetwork}(a) and Fig.\ref{fig:mutualperformancenetwork}(b), two different domain images (``Hidden Factors A'' and ``Hidden Factors B'') are chosen as hidden factors to train the network with the same architecture. The Fig.\ref{fig:mutualperformancenetwork}(c) is original image, Fig.\ref{fig:mutualperformancenetwork}(d) is the image generated by the encrypted network which is trained by ``Hidden Factors A'', and Fig.\ref{fig:mutualperformancenetwork}(e) presents the result of decrypting ciphertext image through the decryption network trained with ``Hidden Factors B''. From the experiment, it can be found that the image generated by the encrypted network which is trained by ``Hidden Factors A'' cannot be decrypted by the network trained by ``Hidden Factors B''.
Therefore, it can be proven that the ``attack network'' with the same architecture trained by different hidden factors, cannot be used to decrypt the ciphertex image with each other. That is, even if attackers obtain the network architecture, they cannot train the decryption network to decrypt the encrypted image without knowing the hidden factors.

\subsubsection{Both Hidden Factors and Network Architecture Leakage}
In this experiment, the network is trained with four times under the same hidden factors and training conditions to get the networks A, B, C and D, respectively.
The experiment evaluates the decryption performance for these four networks on the same ciphertext image to verify whether the parameters generated for each network are different. As shown in Fig.11, the gray value distribution of the image decrypted by the decryption key B, the decryption key C, and the decryption key D is completely different from the image decrypted by the decryption key A. It can be clearly found that under the same training condition, the encrypted medical image encrypted by one network, cannot be decrypted by adopting the parameters in other network. Even if the model parameters are trained with the same network architecture and the same hidden factors, they cannot be used to decrypt the image with each other. Experiments show that even if both the network architecture and the hidden factors are leaked, and training the network under the same training conditions, the parameters of each network are totally different, i.e., the secure keys are different.
It can be proven that DeepEDN is secure even if the network architecture and hidden factors are revealed.
\begin{figure}[t]
\centerline{\includegraphics[width=0.5\textwidth]{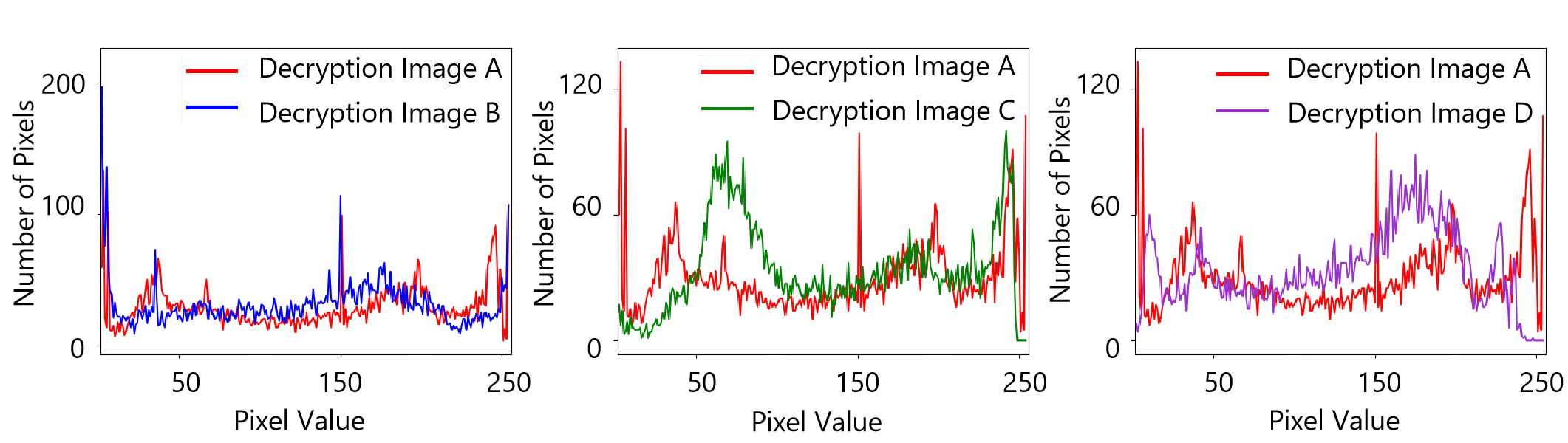}}
\caption{The decryption performance for these four networks on the same ciphertext image.}
\label{fig:decryptionperformance}
\end{figure}

\subsection{Security Analysis under Different Attack Models}
\subsubsection{Ciphertext Only Attack}
In this type of attack, the attacker has access to a string of ciphertext, but cannot access to the corresponding plaintext.

In DeepEDN, the key space of the encryption model can be expressed as $(2^{32})^{2757936}$ and it is very hard for attacker to break down. At the same time, the privacy key generated with multiple iterations and diffusions is complex. Therefore, it is difficult to crack the ciphertext through ciphertext only attacks.

\subsubsection{Known Plaintext Attack}
The known plaintext attack means that the attacker knows a string of plaintext and the corresponding ciphertext. The attacker will try to decrypt the rest of the ciphertext by using these known information.

In traditional sequential pixel visiting pattern methods, concrete encryption factors, which are generally retrieved as equivalent keys, can be used to recover the received ciphertexts. Taking XOR encryption as an example, the masks calculated directly from plaintext and ciphertext are sufficient to decode the ciphertext.
Typically, masks sequentially correspond to the plain pixels and the retrieved masks by plaintext attack can be directly adopted to crack other ciphertexts.
However, the proposed algorithm adopted the non-sequential encryption mechanism. Without the knowledge of the pixel visiting pattern, the privacy key cannot be obtained by the attacker, thus making plaintext attack infeasible. The proposed algorithm adopts the iteration and diffusion procedures to generate the privacy key. These kinds of producers can significantly improve the security performance and provide additional immunity of the cipher to against known plaintext attack.

\subsubsection{Chosen Plaintext Attack}
In this type of attack, the attacker can access the encryption device, choose a string of plaintext and construct its corresponding ciphertext string.

Generally, an attacker can observe the change of the ciphertext image by making small changes to the plaintext image, such as changing the value of only one pixel of the ciphertext, so as to obtain the connection between the plaintext image and the ciphertext image. This type of attack is called as differential attack which is a kind of chosen plaintext attack method. If a small change in the plaintext image can cause a huge change in the ciphertext image, this differential attack method usually fails to take effect. It indicates that the encryption algorithm can resist this chosen plaintext attack method. Here, the Number of Pixel Change Rate (NPCR)  is adopted to measure the degree of image changing. NPCR refers to the rate of pixels change which indicates the ratio of different pixel values at the same position between two plaintext/ciphertext images. The definition of NPCR is as follows:
\begin{equation}
NPCR = \frac{\sum_{i=0}^W \sum_{j=0}^H D(i, j)}{W \times H}
\label{eq}
\end{equation}
where $W$ and $H$ represent the width and height of the image, respectively. T1 and T2 represent a ciphertext image obtained by encrypting two different plaintext images, respectively.
If $T_1(i,j) = T_2(i,j)$, $D(i, j) = 1$.
If $T_1(i,j) \neq T_2(i,j)$, $D(i, j) = 0$.
In the experiment, there is only about $1\%$ different pixels between these two plaintext images.
Both the original plaintext image and the plaintext image with $1\%$ pixel value changed are input to the proposed encryption model. Then, the NPCR is used to compare the differences between these two encrypted images.
The calculated average NPCR value is $94.21\%$, which means that the information of the plaintext image is well diffused into the ciphertext image.
Since DeepEDN has good diffusion performance and is with highly sensitive to the plaintext, it achieves a good performance to resist the chosen plaintext attack like the differential attack.

\begin{figure*}[htbp]
\centerline{\includegraphics[width=0.95\textwidth]{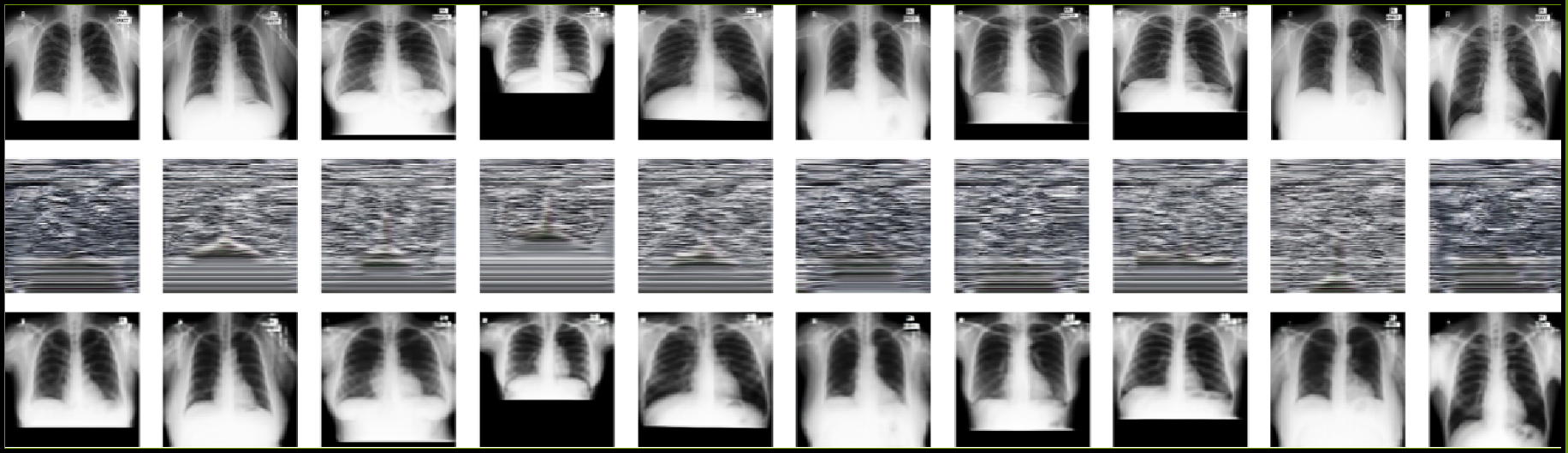}}
\caption{The encryption and decryption performance of the proposed method, in which, the images are original image, encryption image and decryption image from the top to bottom, respectively.}
\label{fig:encryptionanddecryption}
\end{figure*}

\subsubsection{Chosen Ciphertext Attack}
In this type of attack, the attacker can access the decryption device, choose a string of ciphertext and construct its corresponding plaintext string.

Since the structure of our decryption model is exactly the same as the encryption model, the experiment for chosen ciphertext attack is similar to that in chosen plaintext attack. In this experiment, the input of the decryption network is the ciphertext image and the NPCR is used to calculate the different between two decrypted images. According to the experiment, it is found that when the input ciphertext image changes slightly ( just $1\%$ pixels changed), the average NPCR value between two decrypted images is $94.87\%$. It means that if the input ciphertext image changes slightly, the decrypted image will change dramatically. This demonstrates that the proposed algorithm has good diffusion performance and is also highly sensitive to the ciphertext. It is  effective to resist the chosen ciphertext attack.

\section{Experimental Results} \label{sec:experiment}
\subsection{Performance of Encryption and Decryption}
As shown in Fig.\ref{fig:encryptionanddecryption}, the results of the proposed method for medical image encryption and decryption are presented in a visual way. It can be seen that the ciphertext image generated by the encryption network G, is totally different from the original medical image and the pathology information cannot be observed. In addition, the image in the third row is decrypted from the encrypted one through the decryption network F, can recover the detailed information of the original image and restore to the original one.
In order to evaluate the effectiveness of the decryption network, the Peak Signal to Noise Ratio (PSNR) and structural similarity index (SSIM) are employed as evaluation metrics.

The quantitative measure of the decryption error is PSNR, which is based on the root mean square error (RMSE) between the decrypted data and ground truth. It can be represented as:
\begin{equation}
PSNR = 20log_{10}\frac{255}{RMSE}\label{eq}
\end{equation}
To further evaluate the performance of encryption and decryption, the SSIM is used as another metric.
\begin{equation}
SSIM(x, y) = [l(x, y)]^{\alpha}[c(x, y)]^{\beta}[s(x, y)]^{\gamma}\label{eq}
\end{equation}
where $l(x, y)$ is the brightness comparison, $c(x, y)$ is the contrast comparison, and $s(x, y)$ is the structure comparison. The closer the SSIM is to 1, the more resemblance the two images are. And if this value approaches to 0, the two images are completely different. In an ideal case, the SSIM between the encrypted image and the original image is equal to 0, and the SSIM between the decrypted image and the original image is equal to 1. As shown in the second and third rows of Table VI, the SSIM between the encrypted image and the original image is close to 0, and the SSIM between the decrypted image and the original image is close to 1.

For most of medical image processing tasks, the image can be compressed to one-half size of the original one to reduce the storage consumption and does not affect the doctor's diagnosis. In order to ensure that the decrypted image do not affect the doctor's diagnosis, the performance of the reconstructed image decrypted by the decryption network, is also compared with the one-half compressed image. According to the experiment, it is demonstrated that the performance of the reconstructed image is equivalent to that through directly compressing the original image to half and then restoring it. In Table \ref{tab06}, from line 3 to line 6, 2X means that the original image is compressed to one-half and then restored. At this level, human can accurately identify the patient's organ contours and bone information from reconstructed images.
\begin{table*}[htbp]
\caption{Evaluation of the SSIM and PSNR.}
\begin{center}
\begin{tabular}{|c|c|c|c|c|c|c|c|c|c|c|}
\hline
\textbf{Image Id} &\textbf{1} &\textbf{2} &\textbf{3} &\textbf{4} &\textbf{5} &\textbf{6} &\textbf{7} &\textbf{8} &\textbf{9} &\textbf{10} \\
\hline
SSIM(Encrypted) &0.93 &0.88 &0.90 &0.94 &0.93 &0.91 &0.91 &0.93 &0.91 &0.89 \\
\hline
SSIM(Decrypted) &0.01 &0.02 &0.01 &0.01 &0.02 &0.02 &0.01 &0.02 &0.01 &0.01 \\
\hline
SSIM(2X) &0.90 &0.92 &0.90 &0.92 &0.89 &0.91 &0.88 &0.90 &0.91 &0.90 \\
\hline
PSNR &37.43 &35.34 &36.01 &38.03 &35.76 &35.87 &36.13 &37.17 &35.88 &35.74 \\
\hline
PSNR(2X) &35.48 &35.74 &35.03 &35.28 &34.87 &36.73 &34.75 &34.61 &36.17 &34.80 \\
\hline
\end{tabular}
\label{tab06}
\end{center}
\end{table*}
\subsection{Performance of ROI-Mining-Network}
\begin{figure}[t]
\centerline{\includegraphics[width=0.48\textwidth]{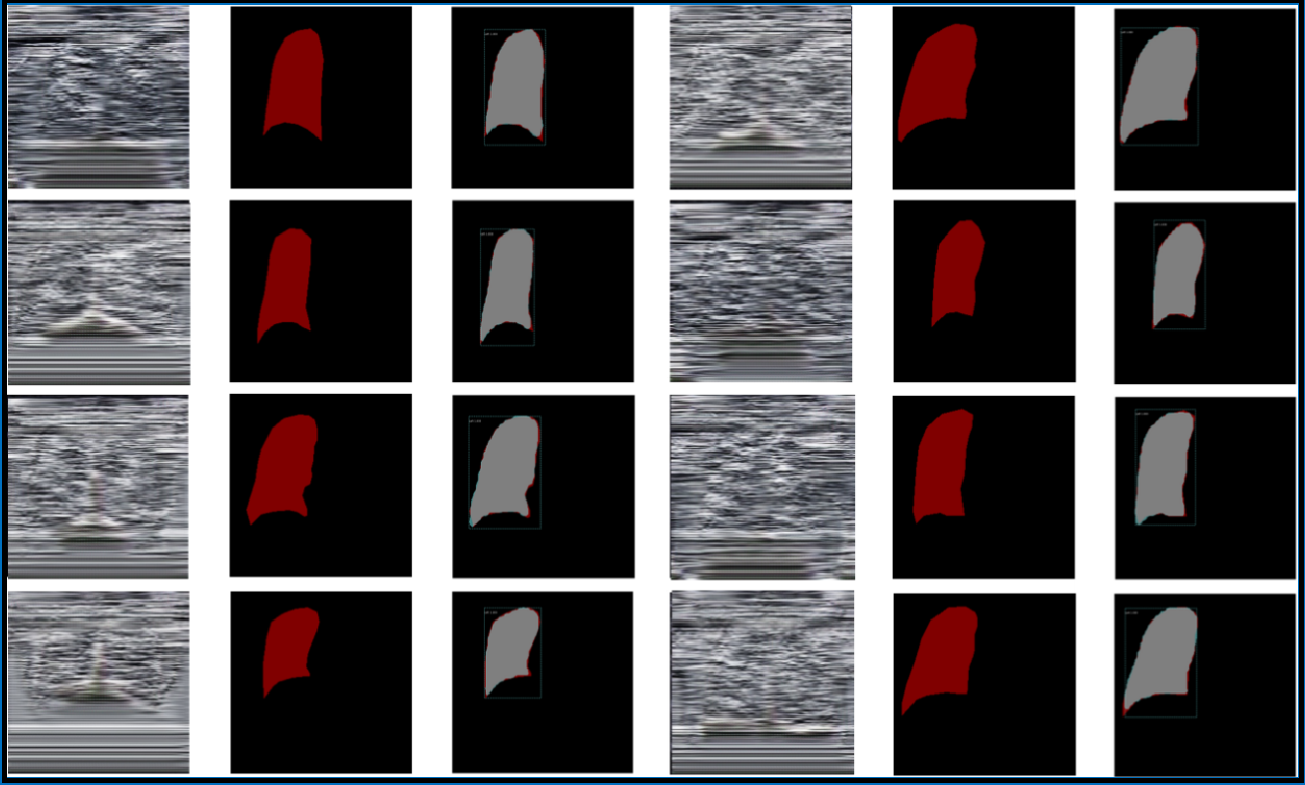}}
\caption{The performance of ROI-Mining-Network.}
\label{fig:performanceofROI}
\end{figure}

Direct extraction of interested information under ciphertext conditions is of great significance for medical image security and also for the data mining with privacy protection. The proposed ROI-mining-network can segment the patient's interested organ tissue from the ciphertext image without decrypting the image firstly. The proposed network has the ability to realize the data mining from the privacy environment by extracting the ROI from the encrypted image directly.  In order to evaluate the proposed ROI-mining-network, the well-known evaluation metric Dice score is adopted in here and is defined as:
\begin{equation}
Dice(GT, AT) = \frac{GT \cap AT}{(|GT|+|AT|)/2}\label{eq}
\end{equation}
The GT represents the ground truth and the AT represents the model predictions. Fig.\ref{fig:performanceofROI} shows the performance of the proposed ROI-mining-network on the patient's left lung. It can be clearly seen that the prediction (grey ones) obtained from the model is almost as the same as the ground truth (red ones).
In addition, the original medical images are also adopted as the experiment data for training the same ROI-mining-network, which is mainly used as the comparison. Under the same training conditions, the DICE of the segmentation network for plaintext is 0.967, while the DICE of the segmentation network for the ciphertext image is 0.962. It can be proven that the ROI-mining-network can achieve a good segmentation performance on both plaintext and ciphertext images.

\begin{figure}[t]
\centerline{\includegraphics[width=0.48\textwidth]{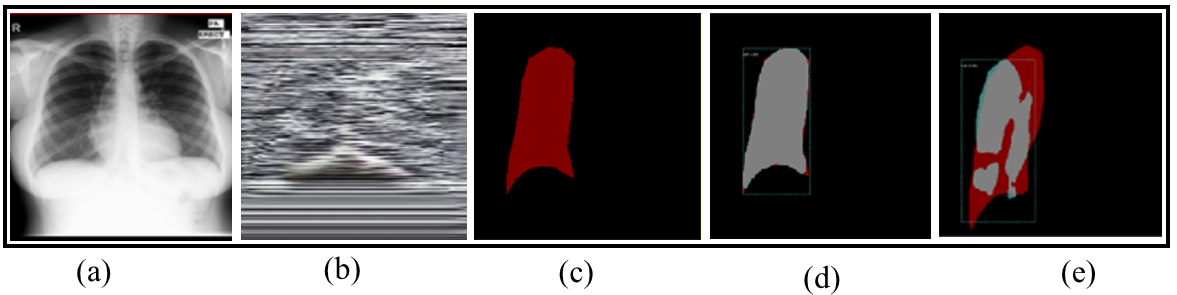}}
\caption{Attack experiment for the proposed ROI-Mining-Network.}
\label{fig:attackforROI}
\end{figure}

As mentioned before, the privacy keys of the network are totally different when training the network at different times even if all the conditions are the same.
Therefore, the attacker cannot obtain the same ROI-mining-network even if employing the same ciphertext image for training. The experiment can be found in Fig.\ref{fig:attackforROI}. In this experiment, Fig.\ref{fig:attackforROI} (a) is the original image and Fig.\ref{fig:attackforROI} (b) is the ciphertext image of Fig.\ref{fig:attackforROI} (a). Fig.\ref{fig:attackforROI} (c) is the ground truth for the right lung segmentation. Fig.\ref{fig:attackforROI} (d) is the correct extraction result segmented by the ROI-mining-network. Fig.\ref{fig:attackforROI} (e) is the error extraction result segmented by the attacker.
\subsection{Efficiency}
To evaluate the efficiency of the proposed network, the running speed of encryption and decryption process on different resolution medical images is evaluated. For 256*256 resolution, the proposed network can encrypt or decrypt 14.28 medical images  per second, while the speed is 3.65 images/second for encrypting or decrypting 512*512 resolution image. This encryption/decryption speed can basically meet the efficiency requirement in clinical practice. In addition, some chaotic encryption algorithms have been adopted as the comparison method for evaluating the efficiency. For instance, Zhou \emph{et al}\cite{b32} introduce a simple chaotic system, which employs a combination of two existing one-dimension (1D) chaotic maps (seed maps). Liao \emph{et al}\cite{b33} introduce a novel image encryption algorithm based on self-adaptive wave transmission. Wu \emph{et al}\cite{b35} introduce a wheel-switch chaotic system for image encryption. In \cite{b36}, the proposed method firstly adopts the two-dimensional logistic map with complicated basin structures and attractors for image encryption. This method can encrypt an intelligible image into a random-like one both from the point of view of the statistical and the human visual system.

\begin{figure}[!t]
\centerline{\includegraphics[width=0.45\textwidth]{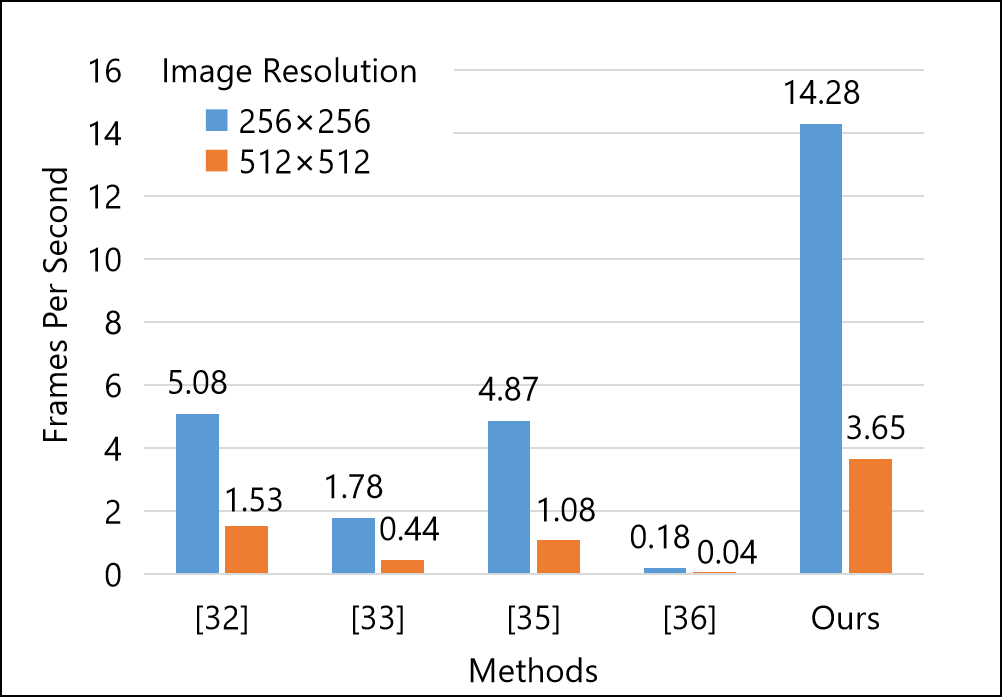}}
\caption{The efficiency comparison between our method and other existing methods.}
\label{fig:efficiency}
\end{figure}

Fig. \ref{fig:efficiency} shows the comparison among aforementioned five chaotic encryption algorithms and the proposed method. The FPS represents the number of images that can be encrypted/decrypted in one second. It can be found that our methods achieve the fastest encryption speed both on 512$\times$512 resolution and 256$\times$256 resolution images. Although the number of keys in our method is greater than the number of keys used in chaotic encryption methods, the processing time of our method is still with higher efficiency.
\section{Conclusion} \label{sec:conclusion}
In this paper, a novel medical image encryption and decryption method (namely DeepEDN) is proposed by leveraging deep learning techniques, which is one of the early attempts to adopt the concept of ``deep learning'' for medical image encryption. The Cycle-GAN network is adopted as the learning network to encrypt and decrypt the medical image. A target domain is used to guide the learning model in the encryption process.
The reconstruction network can decrypt the encrypted image to the original image (plaintext).
Moreover, a ROI-mining-network is proposed to directly extract the ROI from the encrypted medical image, with which DeepEDN can segment the interested organ or tissue in the ciphertext environment without decrypting the medical image. We conduct experiments on the chest X-ray datasets and the results show that the proposed algorithm can protect the medical image with a high security level and can encrypt/decrypt the image in a more efficient way, compared with state-of-the-art medical image encryption methods.



\begin{IEEEbiography}[{\includegraphics[width=1in,height=1.25in,clip,keepaspectratio]{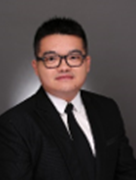}}]
{Yi Ding} received the B.S. degree in software engineering from the University of Electronic Science and Technology of China and the Ph.D. degree in computer science from the Dublin Institute of Technology, Ireland, in 2008 and 2012, respectively. From 2012 to 2016, he was a postdoc research fellow at University of Electronic Science and Technology of China. Since 2014, he has been an associate professor in the School of Information and Software Engineering, University of Electronic Science and Technology of China. He is also a researcher in Institute of Electronic and Information Engineering of UESTC in Guangdong, China. His research interested include machine learning, deep learning, medical image processing, and computer-aided diagnosis.
\end{IEEEbiography}

\begin{IEEEbiography}[{\includegraphics[width=1.1in,height=1.25in,clip,keepaspectratio]{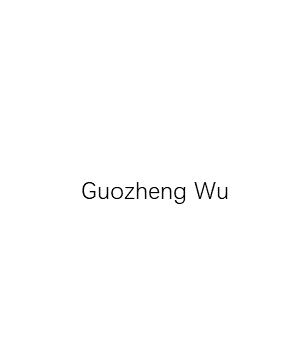}}]
{Guozheng Wu} received the B.S. degree in Information Management and Decision Science from the University of Science and Technology of China and the M.S. degree in Public Management from the Tsinghua University, China, in 1997 and 2002, respectively. He received the Ph.D degree from University of Electronic Science and Technology of China in 2010. He is currently a staff in National Natural Science Foundation of China. His research interested include computer networking, information security, cryptography, information management and software engineering.
\end{IEEEbiography}

\begin{IEEEbiography}[{\includegraphics[width=1in,height=1.25in,clip,keepaspectratio]{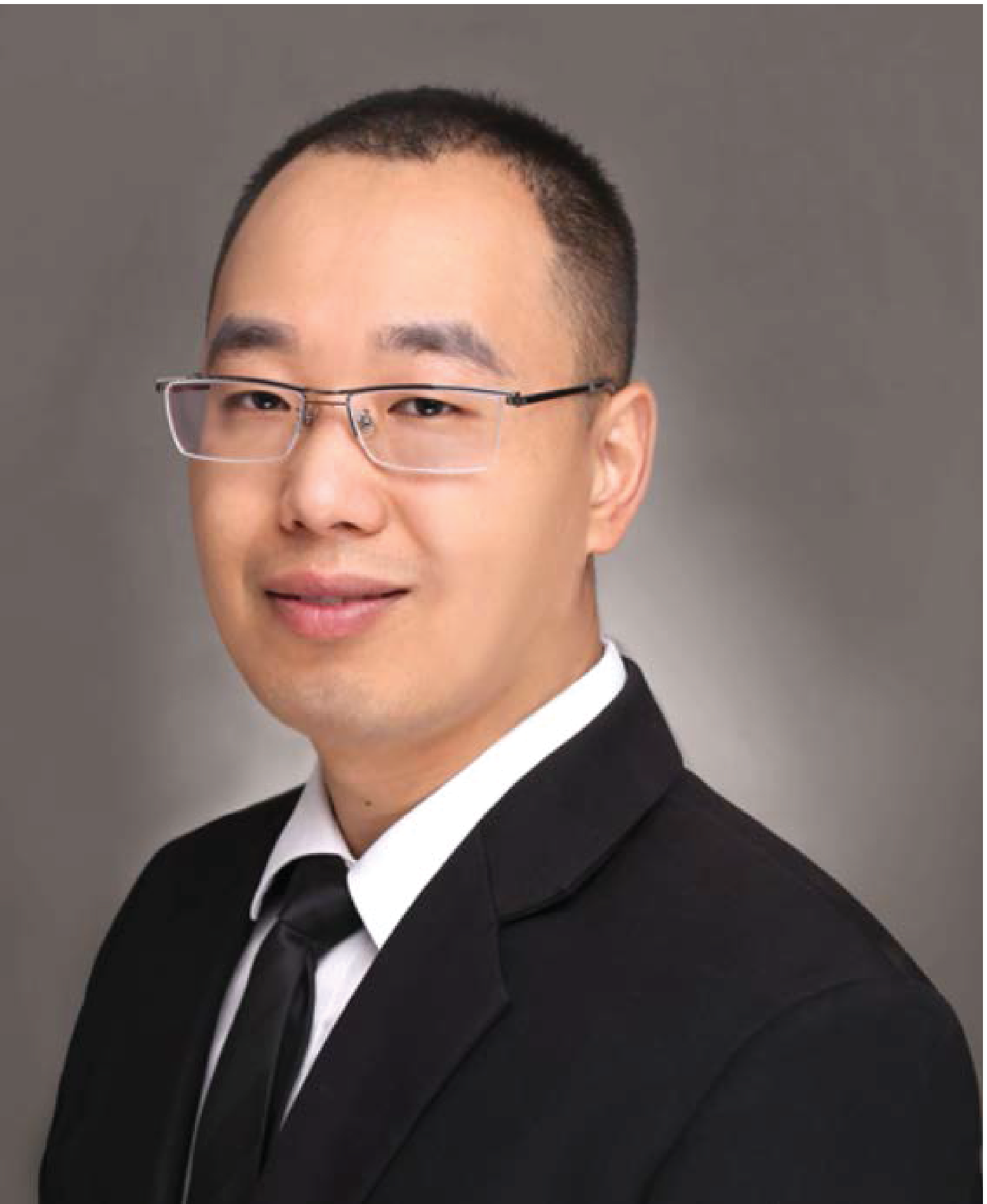}}]{Dajiang Chen} (M'15) is currently an Associate Professor in the School of Information and Software Engineering at University of Electronic Science and Technology of China (UESTC). He was a Post Doctoral Fellow
with the Broadband Communications Research (BBCR) group, Department of Electrical and Computer Engineering, University of Waterloo, Canada, from 2015 to 2017.
He received the Ph.D. degree in information and communication engineering from UESTC in 2014. Dr. Chen served as the workshop chair for BDEC-SmartCity'19
in conjunction with IEEE WiMob 2019 and the organizer for IoT track in conjunction with EAI CollaborateCom 2020. He also serves/served as a Technical Program Committee Member for IEEE Globecom, IEEE ICC, IEEE VTC, IEEE WPMC, and IEEE WF-5G.
His current research interest includes Wireless Security, Physical Layer Security, Secure Channel Coding, and  Machine Learning and its applications in Wireless Network Security and Wireless Communications.
\end{IEEEbiography}

\begin{IEEEbiography}[{\includegraphics[width=1in,height=1.25in,clip,keepaspectratio]{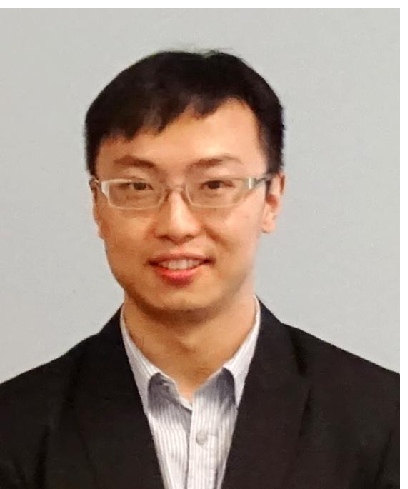}}]
{Ning Zhang} (M'15-SM'18) received B.E. degree
and M.S. degree from Beijing Jiaotong University
and Beijing University of Posts and Telecommunications
in 2007 and 2010, respectively. He received the
Ph.D degree from University of Waterloo, Canada, in
2015. From 2015 to 2017, he was a postdoc research
fellow at University of Waterloo and University
of Toronto, Canada, respectively. Since 2017, He
has been an Assistant Professor at Texas A$\&$M
University-Corpus Christi, USA. He serves/served
as an Associate Editor of IEEE Transactions on
Cognitive Communications and Networking, IEEE Access and IET Communications,
an Area Editor of Encyclopedia of Wireless Networks (Springer) and
Cambridge Scholars. He also served as the workshop chair for MobiEdge'18
(in conjunction with IEEE WiMob 2018) and CoopEdge'18 (in conjunction
with IEEE EDGE 2018), and 5G$\&$NTN'19 (in conjunction with IEEE ICCC
2019). He is a recipient of the Best Paper Awards from IEEE Globecom
in 2014, IEEE WCSP in 2015, Journal of Communications and Information
Networks in 2018, IEEE Technical Committee on Transmission Access and
Optical Systems in 2019, and IEEE ICC in 2019, respectively. His current
research interests include next generation mobile networks, physical layer
security, machine learning, and mobile edge computing.
\end{IEEEbiography}

\begin{IEEEbiography}[{\includegraphics[width=1.1in,height=1.25in,clip,keepaspectratio]{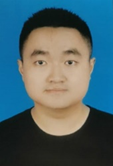}}]
{Linpeng Gong} received B.E. degree from University of Electronic Science and Technology of China in 2016. He is currently pursuing M.S. degree from the University of Electronic Science and Technology of China. His current research interests include deep learning, computer vision, digital image security and Internet of Things.
\end{IEEEbiography}

\begin{IEEEbiography}[{\includegraphics[width=1.1in,height=1.25in,clip,keepaspectratio]{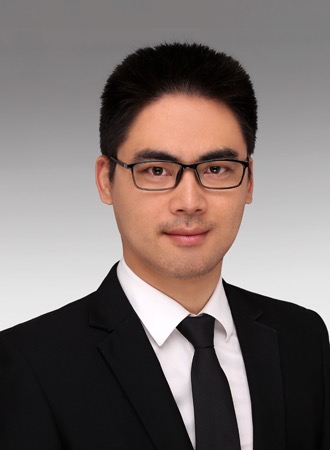}}]
{Mingsheng Cao} currently is a lecturer at Network and Data Security Key Laboratory of Sichuan Province, University of Electronic Science and Technology of China, ChengDu. He received the B.Sc. and M.Sc. degree from UESTC, 2008 and 2011, respectively. In 2019, he obtained the PhD. degree from UESTC. His research interests include Network Security, Pervasive Computing, and  Machine Learning and its applications in Network Security and Pervasive Computing.
\end{IEEEbiography}

\begin{IEEEbiography}[{\includegraphics[width=1in,height=1.25in,clip]{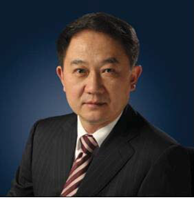}}]{Zhiguang Qin} (S'95-A'96-M'14)
is the Director of the Key Laboratory of New Computer Application Technology and Director of UESTC-IBM Technology Center.
Dr. Qin was the Dean of the School of Software of University of Electronic Science and Technology of China (UESTC).
His research interests include wireless sensor networks, mobile social networks, Information Security, Applied Cryptography, Information Management, Intelligent Traffic, Electronic Commerce, Distribution and middleware, etc.
Dr. Qin served as the General Co-Chair for WASA 2011, Bigcom 2017, etc.
\end{IEEEbiography}

\end{document}